\documentclass[11pt]{article}
\RequirePackage{epsfig}
\RequirePackage{amsmath, esint, amssymb, fancybox, empheq}
\usepackage{xcolor}     
\usepackage{lipsum}     
\usepackage{mdframed}   
\usepackage{comment} 
\usepackage{color}
\definecolor{lightgray}{gray}{0.9}  

\newcommand{\nab}{\mbox{\boldmath $\nabla$}}

\title{Comments on Barker \& Astoul (2021)}
\author{Caroline Terquem \\ Oxford University}
\date{June 2021}

\begin{document}

\maketitle




\begin{abstract}

The tidal evolution of interacting binaries when the orbital period is short compared to the primary star's convective time scale is a problem of long--standing.   Terquem (2021) has argued that, when this temporal ordering scheme is obeyed, 
the rate of energy transfer from tides to convection (denoted $D_R$) is given by the product of the averaged Reynolds stress associated with the tidal  velocity and the mean shear associated with the convective flow.    In a recent response,  Barker \& Astoul (2021, hereafter BA21) claim to show that $D_R$ (in this form) cannot contribute to tidal dissipation.    Their analysis is based on a study of Boussinesq and anelastic models.
 
 Here, we demonstrate that BA21 misidentify the correct term responsible for ener\-gy transfer  between tides and convection.   As a consequence, their anelastic calculations do not prove that the $D_R$  formulation is invalidated as an energy--loss coupling between tides and convection.   BA21 also carry out a calculation in the Boussinesq approximation.   Here, their claim that $D_R$ once again does not contribute is based on boundary conditions that do not apply to any star or planet that radiates energy from its surface,  which is a key dissipational process in the problem we consider.   

\end{abstract}

\section{Derivation of the exchange of kinetic energy between the tide and convection}

In a recent paper, BA21 describe the tidal interaction between stars in a binary system, challenging a proposal made by Terquem (2021, hereafter T21) that this interaction may be described by a novel form for the coupling between tidal and convective velocities.    T21 applies when the orbital period is short compared to the primary star's convective time scale.   BA21 make the assumption that the equilibrium tide is irrotational {\em everywhere} in the volume of the flow.   However,  as shown by Goodman \& Dickson 1998 and Terquem et al. 1998 (see also Bunting et al. 2019), this condition is actually satisfied only when the tidal period $P$ is shorter than the convective timescale $t_{\rm conv}$.   While this condition is met in the envelope of giant planets (where $P \ll t_{\rm conv}$ for the periods of interest), 
it is not satisfied everywhere in a convective zone similar to that of the Sun.   For example, for a 12-day orbital period, $t_{\rm conv} > P$ above 0.88 solar radius (T21).   Therefore,  the study by BA21 is limited only to giant planets or to those stellar binaries with very short orbital periods.  
 
Accordingly, in what follows, we shall consider a model in which the flow is a superposition of two flows which vary with very different timescales: $P$ and $t_{\rm conv}$, satisfying $P \ll t_{\rm conv}$.

\subsection{Reynolds decomposition and conservation equations}

We use the Reynolds decomposition in which the total velocity is written as ${\bf u} \left({\bf r},t \right)= {\bf V} \left({\bf r},t \right)+ {\bf u}' \left({\bf r},t \right)$, where ${\bf V} $ is the velocity  of the slowly varying flow (here taken to be convection) and ${\bf u}'$ is that of the rapidly varying flow with zero mean (the tide).   
Denoting by brackets an
average over an intermediate time $T$ such that  $P  \ll T \ll  t_{\rm conv}$, we have 
$\left< {\bf u}'\left({\bf r},t \right) \right> = {\bf 0}$ and ${\bf V} \left({\bf r},t \right)=\left< {\bf u} \left({\bf r},t \right)\right>$. 
A similar decomposition can be made for the pressure $p \left({\bf r},t \right)= \overline{p} \left({\bf r},t \right)+ p' \left({\bf r},t \right),$ and the mass density $\rho \left({\bf r},t \right)= \overline{\rho} \left({\bf r},t \right)+ \rho' \left({\bf r},t \right),$ 
with $\overline{p} \left({\bf r},t \right)=\left< p  \left({\bf r},t \right) \right>$,  $\overline{\rho} \left({\bf r},t \right) =\left< \rho  \left({\bf r},t \right) \right>$ and
$\left< p'  \left({\bf r},t \right) \right> = \left< \rho'  \left({\bf r},t \right) \right>=0.$
Both  $\overline{\rho}$ and $\overline{p}$ vary on the long timescale $t_{\rm conv}$. We can further write $\overline{\rho} \left({\bf r},t \right)  = \rho_0 \left({\bf r}\right) + \delta \rho \left({\bf r},t \right)$, where $\rho_0 \left({\bf r} \right) = \left< \overline{\rho}  \left({\bf r},t \right) \right>_{\tau}$ is independent of time, with the brackets denoting an average over a time $\tau$ large compared to $t_{\rm conv} $. Similarly, $\overline{p} \left({\bf r},t \right) =p_0 \left({\bf r} \right) + \delta p \left({\bf r},t \right)$.  In other words, $\rho'$ and $p'$ are the zero-mean density and pressure perturbations due to the tide, whereas $\delta \rho$ and $\delta p$ are the zero-mean fluctuations due to convection.   

While the tidal velocity is not a turbulent flow,  it fluctuates much more rapidly with time than the convective velocity (in the regime we are interested in).  Thus,  the equations are in effect the same as that of a turbulent shear flow,  as discussed in detail in T21.   


The net flow satisfies the Navier--Stokes equation, whose Cartesian $i$--component  is:
\begin{equation}
 \rho  \frac{\partial u_i}{\partial
  t} +    \rho \left( {\bf u} \cdot \nab \right) u_i =    -  \frac{\partial }{\partial x_i} \left( \delta p + p' \right) +\left( \delta \rho + \rho' \right)   g_i + f_{t,i},
\label{eq:NSturb}
\end{equation}
where ${g_i}$ is the acceleration due to the stellar gravitational force, and $f_{t,i}$ is the tidal force.   
We have subtracted off the hydrostatic equilibrium equation $\partial p_0 / \partial x_i = \rho_0 g_i$. Viscous forces are neglected for simplicity, but they were take into account in T21.  
In the Boussinesq and anelastic approximations used in this note,  $\left| \delta \rho + \rho' \right|$ will be neglected compared to $\rho_0$ on the left--hand side of 
 this equation.  In other words, the density fluctuations are kept only in the buoyancy term, where they are always multiplied by ${\bf g}$.

Next, we use the Reynolds decomposition,
and average  equation~(\ref{eq:NSturb}) over the intermediate time  $T$.  This  yields: 
\begin{equation}
 \rho_0 \frac{\partial V_i }{\partial
    t} +  \rho_0 \left( {\bf V} \cdot \nab \right) 
      V_i 
  + \rho_0 \left< \left(   {\bf u}' \cdot \nab \right)  u'_i \right>=   -  \frac{\partial \delta p }{\partial x_i}  + \delta \rho   g_i  ,
   \label{eq:NSturb1}
\end{equation}
where we have used $\left< f_{t,i} \right>=0$ since the  the tidal force varies on the short timescale $P$.

The flow also satisfies the mass conservation equation, which can be written as:
\begin{equation}
\frac{\partial }{\partial t}  \left( \delta \rho + \rho' \right) + \nab \cdot \left[ {\rho}_0 \left( {\bf V} + {\bf u'} \right) \right ] =0,
\label{mass}
\end{equation}
where in the divergence term the density fluctuations have been neglected compared to $\rho_0$.

\subsection{Boussinesq approximation}

In this approximation,  the density fluctuations are completely neglected in equation~(\ref{mass}), so that the entire time-derivative term is ignored.  Also,  
$\rho_0 \left({\bf r}\right) = \rho_0$ is assumed to be uniform.   Equation~(\ref{mass}) then implies that both 
the mean and fluctuating flows are incompressible: 
$  \nab \cdot {\bf V}    =
  \nab \cdot {\bf u}'  =0.$    In other words, compressibility is only retained in the buoyancy term in the Navier--Stokes equation.   In principle, this approximation is strictly valid only if the height of the flow is small compared to the scale over which the thermodynamic parameters change (Spiegel \& Veronis 1960), which of course is not satisfied when considering the convective zone of the Sun.  

We obtain an  equation for the kinetic energy of the mean flow per unit volume, $E = \rho_0 V_i V_i / 2$, by multiplying
equation~(\ref{eq:NSturb1}) by $V_i$ and summing over $i$.  Using the incompressibility conditions then yields:
 \begin{equation}
  \frac{\partial  E}{\partial
    t}  =
-  \frac{\partial  }{\partial x_j}  \left(E V_j  + \delta p V_j \right)  
  +  \frac{\partial}{\partial x_j} \left(
R_{ij} V_i \right) +  \rho_0 D_R  + \delta \rho g_i V_i
,
   \label{eq:NSturbE2}
   \end{equation}
where we have noted $R_{ij}=-\rho_0  \left<  u'_j  u'_i  \right> $ the Reynolds stress tensor and have defined
\begin{equation}
 \rho_0 D_R \equiv - R_{ij}  \frac{ \partial V_i }{ \partial x_j } ,
\label{eq:DR}
\end{equation}
and repeated indices are summed over.  A similar conservation equation for the mean kinetic energy of the fluctuations per unit volume, $\left< e'  \right> = \rho_0  \left< u'_i u'_i  \right>/ 2$, can be obtained
by multiplying equation~(\ref{eq:NSturb}) by $u'_i$.  Substituting the Reynolds decomposition, averaging over time and using the incompressibility conditions then yields:
\begin{equation}
   \frac{\partial \left< e' \right> }{\partial
    t}  = -  
  \frac{\partial  }{\partial x_j}  \left( \left< e'  \right>  V_j + \left< p' u'_j  \right> \right) 
    - \frac{\partial}{\partial x_j}
  \left(
   \frac{\rho_0}{2} \left<  u'_j  u'_i u'_i 
   \right>
\right)  - \rho_0 D_R + \left<\rho' g_i  u'_i  \right>  + \left< f_{t,i} u'_i \right>.
   \label{eq:NSturbE5}
   \end{equation}

We now integrate equations (\ref{eq:NSturbE2}) and (\ref{eq:NSturbE5})  over some control volume $v$ bounded by a surface  $s$,  to obtain:
 \begin{empheq}[box=\fbox]{align}
& \frac{{\rm d}  }{{\rm d}
    t} \iiint_{v}   E  {\rm d} v  =   - \oiint_{ s}   \left(  E + \delta p \right) V_j  n_j  {\rm d} s -F   + {\cal D} + W_b ,
    \label{KE1}
    \\
&    \frac{{\rm d}  }{{\rm d}
    t} \iiint_{v}  \left< e' \right>  {\rm d} v =   - \oiint_{s}  \left(  \left< e'  \right>  V_j + \left< p' u'_j  \right> \right) n_j {\rm d} s -  F'  - {\cal D} + W'_b + W_t ,
   \label{KE2}
\end{empheq}
where $n_j$ is the $j$--component of the unit vector normal to the surface, and we have defined:
\begin{equation}
 F =  - \oiint_{s}  R_{ij}  V_i  n_j {\rm d} s , \; \; \;
 F'  =  \oiint_{ s} \frac{\rho_0}{2} \left<  u'_j  u'_i u'_i 
   \right> n_j {\rm d} s , 
\end{equation}
and:
\begin{equation}
{\cal D}  =  \iiint_{v}  \rho_0 D_R {\rm d} v, \; \; \; W_b = \iiint_{v}  \delta \rho g_i V_i  {\rm d} v , \; \; \; W'_b = \iiint_{v}  \left<\rho' g_i  u'_i  \right> {\rm d} v , \; \; \; W'_t =  \iiint_{v}  \left< f_{t,i} u'_i \right>  {\rm d} v.
\end{equation}

BA21 claim that the {\em sum} ${\cal}I_{ee}=-F+{\cal D}$ is the term that exchanges kinetic energy between  the tide and convection.    This is incorrect, as we discuss below by interpreting the different terms in the equations.   

In equation~(\ref{KE1}), the first term on the right--hand side is the transport of mean kinetic energy by the mean flow through the surface $s$, together with the work done by the fluctuations of the pressure associated with convection on the mean flow. 
The second term, $-F$, represents 
the work done by the Reynolds stress on the mean flow.  If  the mean velocity at the surface is parallel to the surface,  $F=0$   and  the $\partial \left( R_{ij} V_i \right) / \partial x_j $ term in equation (\ref{eq:NSturbE2}) only redistributes energy within the volume of the fluid.  The critical third term, ${\cal D}$, will be discussed below. 
The last term, $W_b$, represents the input of kinetic energy in the mean flow from the buoyancy force.   
 
Similarly, in equation~(\ref{KE2}), the first term on the right--hand side describes the transport of mean kinetic energy of the fluctuating velocities by the mean flow through the surface, and the work done by the fluctuations of the pressure associated with the tide on the fluctuating velocities.   The second term, $-F'$, represents the work done by the Reynolds stress on the fluctuations.  The last two terms, $W'_b$ and $W_t$, represent the work done by the buoyancy and  tidal forces on the fluctuations.  
 
The term ${\cal D}$ is known as the {\em deformation work} (Tennekes \& Lumley 1972).  
 It  arises from the fact that the work done by the Reynolds stress on the surfaces of a fluid element not only changes the bulk velocity of this fluid element, it also deforms the volume, in exactly the same way as the viscous stress (not included here) does.   If the volume element were solid, this work would be stored as elastic potential energy.  In a liquid however, it corresponds to an {\em irreversible} exchange of energy between the fluctuations and the mean flow.   The fact that ${\cal D}$ appears with opposite signs in both equations~(\ref{KE1}) and~(\ref{KE2}) makes it clear that this term exchanges kinetic energy between the mean flow and the fluctuations.   
 
 Equations~(\ref{KE1})  shows that $I_{ee}=-F+D$ is the rate at which the kinetic energy of the mean flow is changed due to interaction with the fluctuations.  However, as seen from equation~(\ref{KE2}), this is {\em not} the rate at which the mean kinetic energy of the fluctuations is changed due to  interaction with the mean flow.   {\em The only term through which convection can extract energy from the tide is ${\cal D}$}.
 When $I_{ee}=0$,  and if  $F$ is  non zero,  the energy input into the mean flow due to the deformation work ${\cal D}$ is exactly balanced by the work done by the Reynolds stress on the mean flow.  
 However, in that case, the deformation work ${\cal D}$ still results in a change of kinetic energy for the fluctuations.    
 

\subsection{Anelastic approximation}

Here, $\rho_0$ is no longer considered to be uniform, and therefore the convective and tidal flows are no longer incompressible.
  Averaging equation~(\ref{mass}) over the intermediate time $T$ yields $\partial \left( \delta \rho \right) / \partial t + \rho_0 \nab \cdot  {\bf V} + {\bf V} \cdot \nab \rho_0 =0$.
The first term on the left--hand side of this equation is $\sim \left| \delta \rho \right| / t_{\rm conv}$. The second and third terms are 
$\sim \rho_0 / t_{\rm conv}$, since ${\bf V}$ varies on the same spatial scale $H_{\rho}$  as $\rho_0$, and $H_{\rho}/V \sim t_{\rm conv}$.  Therefore, the first term is by assumption negligible, which yields:
\begin{equation}
\nab \cdot \left( \rho_0 {\bf V} \right)=0.
\label{AN1}
\end{equation}
Substituting this into the mass conservation equation then gives:
\begin{equation}
 \frac{\partial \rho'}{\partial t} + \nab \cdot \left( \rho_0 {\bf u'} \right)=0.
 \label{AN2}
\end{equation}
In the equilibrium tide approximation, $\left| \rho' \right| \sim  \rho_0 \xi_r / H_{\rho}$ (e.g., Terquem et al. 1998), where $\xi_r$ is the radial tidal displacement.  In the equation above, the divergence term  is $\sim \rho_0 \left| u' \right| /H_{\rho}$ (as ${\bf u'}$ varies on a spatial scale $\sim r > H_{\rho}$).
Since $ \left| \xi_r \right| \sim \left| u' \right| P$ and $\left| \partial \rho' / \partial t \right| \sim \left| \rho' \right| /P$,   the time derivative term  is comparable in magnitude to the divergence term, and therefore must be retained.  

We next derive equations for the kinetic energy of the mean flow and the fluctuations in the same way as with the Boussinesq approximation, but replacing  the incompressibility conditions by equations~(\ref{AN1}) and~(\ref{AN2}).   We then obtain:

\begin{equation}
  \frac{\partial  E }{\partial
    t}  =
-  \frac{\partial  }{\partial x_j}  \left(E V_j  + \delta p V_j \right)  
  +  \frac{\partial}{\partial x_j} \left(
R_{ij} V_i \right) + \rho_0 D_R + \delta \rho g_i V_i - \left< u'_i  \frac{\partial \rho'}{\partial t}  \right> V_i  + \delta p \frac{\partial V_j  }{\partial x_j} 
,
   \label{eq:NSturbE2AN}
   \end{equation}
\begin{multline}
   \frac{\partial \left< e' \right> }{\partial
    t}  = -  
  \frac{\partial  }{\partial x_j}  \left( \left< e'  \right>  V_j + \left< p' u'_j  \right> \right) 
    - \frac{\partial}{\partial x_j}
  \left(
   \frac{\rho_0}{2} \left<  u'_j  u'_i u'_i 
   \right>
\right)  - \rho_0 D_R + \left<\rho' g_i  u'_i  \right>  + \left< f_{t,i} u'_i \right> \\ - \frac{1}{2} \left< u'^2_i  \frac{\partial \rho'}{\partial t} \right> + \left< p'  \frac{\partial u'_j  }{\partial x_j} \right> .
   \label{eq:NSturbE5AN}
   \end{multline}
  Here again, we integrate these equations over a control volume $v$ to obtain:
 \begin{empheq}[box=\fbox]{align}
& \frac{{\rm d}  }{{\rm d}
    t} \iiint_{v}   E {\rm d} v  =   - \oiint_{ s}   \left(  E + \delta p \right) V_j  n_j  {\rm d} s -F   + {\cal D}  + W_b -t_1 + {\cal D}_{\delta p} ,
    \label{KE1AN}
    \\
 &   \frac{{\rm d}  }{{\rm d}
    t} \iiint_{v}  \left< e'  \right>  {\rm d} v  =   - \oiint_{s}  \left(  \left< e'  \right>  V_j + \left< p' u'_j  \right> \right) n_j {\rm d} s -  F'  - {\cal D}  + W'_b + W_t  -t'_1 + {\cal D}_{p'} ,
   \label{KE2AN}
\end{empheq}
where we have defined:
\begin{eqnarray}
{t_1}  & = &   \iiint_{v}  \left< u'_i  \frac{\partial \rho'}{\partial t}   \right> V_i  {\rm d} v  ,  \; \; \;
{\cal D}_{\delta p}  =  \iiint_{v}  \delta p \frac{\partial V_j  }{\partial x_j} {\rm d} v , \nonumber \\
{t'_1}  & = &  \iiint_{v}   \frac{1}{2} \left< u'_i u'_i  \frac{\partial \rho'}{\partial t}  \right> {\rm d} v ,  \; \; \;
{\cal D}_{p'}  =  \iiint_{v} \left< p'  \frac{\partial u'_j  }{\partial x_j} \right>  {\rm d} v .
\end{eqnarray}
Compared to equations (\ref{KE1}) and (\ref{KE2}), there are two extra terms on the right--hand side of equations (\ref{KE1AN}) and (\ref{KE2AN}).   The terms ${\cal D}_{\delta p}$ and ${\cal D}_{p'}$ are the deformation work due to  pressure: in a compressible flow, pressure forces, like the Reynolds stress, deform fluid elements.   The terms $t_1$ and $t'_1$ are related to the rate of change of momentum produced by the time--dependence of $\rho'$.  

 BA21 claim that the rate at which energy  is exchanged between fluctuations and mean flow is $I_{ee}=-F+{\cal D} -t_1$.   Again, this is incorrect.   We have already discussed that $F$ appears only in the equation for the kinetic energy of the mean flow, and the same applies to $t_1$: this term contributes to the change of kinetic energy of the mean flow, but it does not contribute to  that of the fluctuations.  As can be seen from the equations above,  apart from the surface terms (which are zero in BA21), the only coupling between the tide and convection which changes the kinetic energy of the tide is the deformation work ${\cal D}$.  If $\delta \rho$ had not been neglected, there would be an extra term $- (1/2) \left< u'_i u'_i \right> \partial ( \delta \rho ) / \partial t$ on the right--hand side of equation~(\ref{eq:NSturbE5AN}), which would also couple mean flow and fluctuations.  However, in the anelastic approximation, this term is of smaller order compared to $\rho_0 D_R$, so that these terms cannot balance each other.  
Therefore,  here again,  {\em the only term through which convection can extract energy from the tide is ${\cal D}$}.

\section{Comments on the BA21 conditions that yield ${\cal D}=0$  }

In the Boussinesq approximation, BA21 obtain ${\cal D}=0$ by assuming that 
either 
(1)  
${\bf u'} \cdot {\bf n}={\bf V} \cdot {\bf n}=0$ along the bounding surface, where ${\bf n}$ is  the vector normal to the distorted surface, or
(2) 
${\bf V}=0$ everywhere on the bounding surface. \\

Let us examine these conditions in more detail:

\begin{itemize}
\item[$\bullet$] The condition ${\bf u'} \cdot {\bf n}=0$  is generally {\em incompatible with an object which radiates as a blackbody} (or with appropriate surface radiative boundary conditions).  When solving the stellar oscillation equations, the proper outer boundary conditions are i) that the surface is free, so that the Lagrangian variation of the pressure is zero, and  ii) the surface radiates as a blackbody,  which gives a relation between the Lagrangian changes in temperature and flux.  The point here is that the tidal displacement of the surface cannot be specified arbitrarily: it must adjust to ensure that these two surface conditions be satisfied, and this requirement in fact produces a nonzero component along the normal to the (distorted) surface.   It is {\it this} component which is responsible for the flux variation associated with stellar oscillations (Dziembowski 1977, Burkart et al. 2012, Bunting \& Terquem 2021).    
  \item[$\bullet$]  Assuming the impenetrability condition, ${\bf V} \cdot {\bf n}=0$, or the more restrictive no--slip condition, ${\bf V} =0$, 
 along  the bounding surface implies that there is no flux of tidal kinetic energy through the (upper or lower) surface of the convective zone.   Within the Boussinesq approximation, this means that no energy of any kind can be transported into or out of the convective zone.   The impenetrability or no--slip boundary conditions make physical sense  only if they are supplemented by an equation which enables the radiative heat flux to take over the transport of energy when the convective fluxes of enthalpy and kinetic energy diminish near the surfaces.   In fact, the sole emphasis on the kinetic energy flux is misplaced.   Kinetic energy transferred by the tide to the convective flow  becomes part of the overall kinetic energy of convection, and can be converted into thermal  energy via pressure acting on eddies which expand or contract.  This thermal energy  is then transported towards the surface of the Sun by the enthalpy flux  (Miesch 2005).  Then, as fluid elements move up, more and more of the thermal energy they contain is transported away by photons.   Therefore, even though there is no direct transport of kinetic energy through the surface of the convective zone, the tidal energy which is {\em initially} kinetic is transferred to the convective flow and escapes along with the rest of the energy that is already present in the convective zone.   This whole complex process, which controls the transport of energy in the Sun, can only be captured by solving the {\em full energy equation, including both kinetic and thermal energies}, not just the kinetic energy equation in isolation, as done by  BA21.
 \end{itemize}
 
 The numerical simulations performed by BA21 using the Boussinesq approximation 
 and these artificial boundary conditions give the results predicted by their analysis.  In the simulations, ${\bf V}=0$  at the surface and the tide is restricted to be irrotational everywhere.  This set--up inevitably leads to the integral of $\rho D_R$ vanishing over the domain of the flow.
The simulations, by construction, can only confirm the unsurprising result that an irrotational tide cannot exchange energy with an incompressible flow enclosed within rigid boundaries.  This is, however, not relevant to stars or planets.    \\
 
 In the anelastic simulations,  BA21 find that ${\cal D} \ne 0$.  They go on to claim that there is no energy exchange between the tide and convection on the basis of $I_{ee}=-F+{\cal D}-t_1=0$.    This result is obtained only for the unrealistic boundary conditions discussed above, but more to the point (and as shown in the previous section), $I_{ee}$ is not the correct term responsible for energy exchange between the tide and convection.   The fact that ${\cal D} \ne 0$ in this calculation would actually imply that kinetic energy {\em is} transferred between the tide and convection! \\

Finally, we emphasise that the standard term $D^{\rm st}_R$, which couples the Reynolds stress associated with  the components of the convective velocity and the shear associated with the tide, {\em is not an alternative} to $D_R$.  When $P \ll  t_{\rm conv}$,  $\left< D^{\rm st}_R \right>=0$ and the only term through which convection may extract energy from the tide is $D_R$.

\section*{Acknowledgements}
I thank Steven Balbus and John Papaloizou for very informative and stimulating discussions,  for their encouragement and for feedback on this note.

\end{document}